\documentclass[aps, prb, twocolumn, amssymb, amsmath, showpacs, superscriptaddress]{revtex4-1}
\usepackage{bm}
\usepackage{times}
\usepackage{graphicx}
\usepackage{color}
\usepackage{dcolumn}
\usepackage[colorlinks=true, letterpaper=true, pdfstartview=FitV, linkcolor=blue, citecolor=blue, urlcolor=blue]{hyperref}
\usepackage{appendix}
\usepackage[normalem]{ulem}

\newcommand{\vect}[1]{\boldsymbol{#1}}                 
\usepackage{helvet}
\usepackage{pifont}
\newcommand{\cmark}{\ding{51}}
\newcommand{\xmark}{\ding{55}}
\usepackage{multirow}
\usepackage{siunitx}
\usepackage{makecell}

\usepackage{colortbl}
\usepackage{xcolor}
\definecolor{lightgreen}{rgb}{0.88, 1.0, 0.88}

\begin{document}

\title{Magnetoelectric Bulk Photovoltaic Effect in Centrosymmetric Solids}
\author{Longjun Xiang}
\thanks{These authors contributed equally to this work.}
\affiliation{College of Physics and Optoelectronic Engineering, Shenzhen University, Shenzhen 518060, China}
\author{Fuming Xu}
\thanks{These authors contributed equally to this work.}
\affiliation{College of Physics and Optoelectronic Engineering, Shenzhen University, Shenzhen 518060, China}
\author{Lei Zhang}
\email{zhanglei@sxu.edu.cn}
\affiliation{State Key Laboratory of Quantum Optics Technologies and Devices, Institute of Laser Spectroscopy, Shanxi University, Taiyuan 030006, China}
\affiliation{Collaborative Innovation Center of Extreme Optics, Shanxi University, Taiyuan 030006, China}
\author{Jian Wang}
\email{jianwang@hku.hk}
\affiliation{College of Physics and Optoelectronic Engineering, Shenzhen University, Shenzhen 518060, China}
\affiliation{Quantum Science Center of Guangdong-Hongkong-Macao Greater Bay Area, Shenzhen 518045, China}
\affiliation{Department of Physics, The University of Hong Kong, Pokfulam Road, Hong Kong, China}


\begin{abstract}
It is commonly believed that a bulk photovoltaic current requires broken spatial inversion symmetry.
However, in this work,
we demonstrate that dc photocurrents can arise in centrosymmetric solids from the bilinear coupling between the electric and magnetic components of light.
This magnetoelectric bulk photovoltaic effect (MBPE) has a quantum-geometric origin
in the dipole moments of the Zeeman quantum metric and Zeeman Berry curvature.
We identify four distinct MBPE mechanisms,
including time-reversal-even and time-reversal-odd injection-like and shift-like photocurrents driven by linearly or circularly polarized light.
Our symmetry analysis shows that the MBPE is allowed in broad classes of two-dimensional centrosymmetric nonmagnetic and magnetic materials
under normally incident light.
As representative examples, we predict sizable MBPEs in centrosymmetric monolayer Pt$_2$HgSe$_3$ and altermagnetic RuF$_4$,
with current densities on the order of $1~\mathrm{mA/m}$.
Notably, the injection-like photocurrent in RuF$_4$ reverses sign upon switching either the light helicity or the N\'eel vector.
Our results extend the bulk photovoltaic effect to centrosymmetric solids
and open new opportunities for electrical readout of altermagnetic order.
\end{abstract}

\maketitle

\noindent \textit{\textcolor{blue}{Introduction.}}---The bulk photovoltaic effect (BPE),
which converts light into a dc photocurrent in homogeneous solids,
represents a fundamental nonlinear optical phenomenon~\cite{Fridkin2001} with potential applications
in energy conversion~\cite{Alexe2011,Bhatnagar2013,Spanier2016} and photodetection~\cite{Osterhoudt2019,Ma2019,Han2021},
and can reveal the geometry and topology of quantum materials~\cite{Morimoto2016,deJuan2017,Ahn2020,Ahn2022,Ma2021,Ma2023}.
Its microscopic origin is commonly attributed to the electric-dipole coupling between light and electrons,
giving rise to the canonical shift and injection photocurrents~\cite{vonBaltz1981,Aversa1995,Sipe2000,Young2012,Wang2019,Dai2023}.
Because the electric field is odd under spatial inversion ($\mathcal{P}$),
the corresponding second-order photocurrent is forbidden in centrosymmetric crystals~\cite{Morimoto2016,deJuan2017,Ahn2020,Ahn2022,Ma2021,Ma2023,
vonBaltz1981,Aversa1995,Sipe2000,Young2012,Wang2019,Dai2023}.
Consequently, broken $\mathcal{P}$ symmetry is required for the conventional electric-dipole BPE,
although alternative photocurrent mechanisms have been proposed in centrosymmetric systems~\cite{Shi2021,Ikeda2023,Sivianes2025}.

The BPE in magnetic materials has also attracted considerable attention,
with broken time-reversal $(\mathcal{T})$ symmetry enabling magnetic shift and injection photocurrents~\cite{Zhang2019,Wang2020,Dong2026LayerPV}.
Separately, the BPE has been studied~\cite{Ivchenko1984,Burger2020,Hornung2021} in the presence of a static external magnetic field.
These developments highlight the important role of magnetism in the BPE.
However, in both cases the optical driving is still mediated by the electric component of light,
with magnetism entering through either intrinsic magnetic order or an external static field.
Light itself also carries an oscillating magnetic component~\cite{EMbook} that couples to electron spin through the Zeeman interaction.
Whether the resulting electric--magnetic bilinear coupling can generate a bulk photovoltaic current,
and what microscopic mechanisms govern such a response, remain largely unexplored.

In this Letter, we establish the magnetoelectric bulk photovoltaic effect (MBPE),
a dc photocurrent generated by the bilinear coupling between the electric and magnetic components of light.
By developing a microscopic quantum theory, we identify four distinct MBPE mechanisms,
including $\mathcal{T}$-even and $\mathcal{T}$-odd injection-like and shift-like photocurrents driven by linearly or circularly polarized light.
Their quantum-geometric origins are rooted in the dipole moments of the Zeeman quantum metric and Zeeman Berry curvature~\cite{xiangZeeman,
xiangSpin, xiangMap}.
Remarkably, our symmetry analysis shows that the MBPE is allowed in broad classes of two-dimensional centrosymmetric nonmagnetic and magnetic materials
under normally incident light.
As representative examples, we predict sizable MBPEs in monolayer Pt$_2$HgSe$_3$~\cite{Marrazzo2018} and altermagnetic RuF$_4$~\cite{Sodequist2024},
with current densities on the order of $1\,\mathrm{mA/m}$.
Notably, the injection-like photocurrent in RuF$_4$ reverses sign upon switching either the light helicity or the N\'eel vector,
enabling electrical readout of the altermagnetic order~\cite{AM2, AM3, AM4}.

\begin{figure}[t!]
\includegraphics[width=0.90\columnwidth]{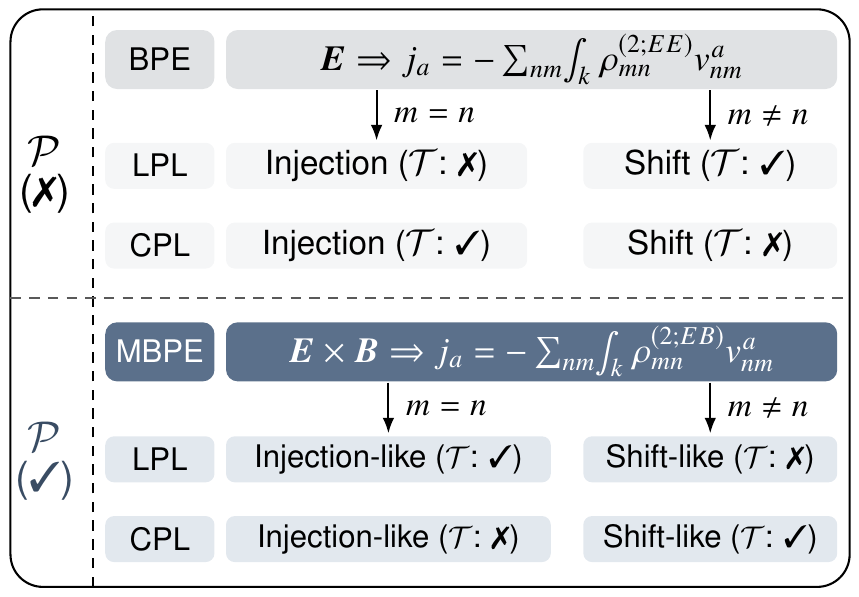}
\caption{\label{FIG1}Bulk photovoltaic effect (BPE) versus magnetoelectric bulk photovoltaic effect (MBPE).
Here $\mathcal{P}$ denotes inversion symmetry, $\mathcal{T}$ time-reversal symmetry, LPL linearly polarized light,
and CPL circularly polarized light.
In addition, \cmark\, (\xmark)\, denotes the presence (absence) of $\mathcal{P}$ and $\mathcal{T}$ in solids.}
\end{figure}

\bigskip
\noindent{\textit{\textcolor{blue}{Theory of the magnetoelectric bulk photovoltaic effect.}}}---We consider
a crystalline solid irradiated by a monochromatic optical field,
whose electric and magnetic components are written as
$\vect{E}(t)=\sum_{\alpha}\vect{E}^{\alpha}(t)=\sum_{\omega_\alpha=\pm\omega}\vect{E}(\omega_\alpha) e^{-i\bar{\omega}_{\alpha}t}$
and
$\vect{B}(t)=\sum_{\alpha}\vect{B}^{\alpha}(t) =\sum_{\omega_\alpha=\pm\omega}\vect{B}(\omega_\alpha) e^{-i\bar{\omega}_{\alpha}t}$,
respectively, where $\bar{\omega}_{\alpha}=\omega_{\alpha}+i\eta$,
and $\eta\rightarrow0^{+}$ implements the adiabatic switching-on of the optical field.
The optical perturbation, including both the electric-dipole and Zeeman couplings,
is given by $\hat{V}(t)=\vect{E}(t)\cdot\hat{\vect{r}} +\bar{\vect{B}}(t)\cdot\hat{\boldsymbol{\sigma}}$,
with $\bar{\vect{B}}(t)=-g\mu_B\vect{B}(t)/2$.
The conventional BPE is driven solely by the electric component of the optical field
and is described by the second-order dc current response~\cite{vonBaltz1981,Aversa1995,Sipe2000,Young2012,Wang2019,Dai2023}
$j^a=\sigma^{abc}(0;\omega,-\omega)E_b(\omega)E_c(-\omega)$.
In contrast, the MBPE is jointly driven by the electric and magnetic components of the optical field
and is described by a bilinear dc current response $j^a=\kappa^{abc}(0;\omega,-\omega) E_b(\omega)\bar{B}_c(-\omega)$.
We set $e=\hbar=1$ unless otherwise stated.

A key distinction between the BPE and MBPE lies in their parity under $\mathcal{P}$.
Since $\vect{j}$ and $\vect{E}$ are polar vectors,
they reverse sign under spatial inversion,
whereas $\vect{B}$, being an axial vector, remains invariant.
As a result, in centrosymmetric solids,
the conventional BPE with a $\mathcal{P}$-odd susceptibility $\sigma^{abc}$ is forbidden,
whereas the MBPE with a $\mathcal{P}$-even susceptibility $\kappa^{abc}$ can remain finite.
Because the Zeeman coupling is typically weaker than the electric-dipole coupling,
the MBPE is naturally expected to be weaker than the conventional BPE and may therefore have been overlooked.
As shown below, however, its magnitude depends strongly on the driving frequency and can reach experimentally measurable values.

Within the density-matrix framework~\cite{vonBaltz1981,Aversa1995,Sipe2000},
the current is given by $j^a=-\text{Tr}[\hat{v}^a\hat{\rho}]$,
where $\hat{v}^a$ and $\hat{\rho}$ are the velocity and density-matrix operators, respectively.
In the Bloch basis, with $\hat{H}_0|n\rangle=\epsilon_n|n\rangle$,
the current becomes $j^a=-\sum_{nm}\int_k \rho_{mn}v^a_{nm}$,
where $\int_k \equiv \int d^dk/(2\pi)^d$, $v^a_{nm}=\langle n|\hat{v}^a|m\rangle$, and $\rho_{mn}=\langle m|\hat{\rho}|n\rangle$.
The density-matrix elements $\rho_{mn}$ are obtained by solving the quantum Liouville equation
$i\partial_t\hat{\rho}=[\hat{H}_0+\hat{V}(t),\hat{\rho}]$ in the Bloch basis.
For the conventional BPE, the susceptibility $\sigma^{abc}$ is determined by the second-order density-matrix element $\rho_{mn}^{(2;EE)}$.
Its diagonal and off-diagonal components give the injection and shift currents~\cite{Sipe2000}, respectively,
each of which can be decomposed into $\mathcal{T}$-even and $\mathcal{T}$-odd contributions~\cite{Wang2020} [upper panel of Fig.~\ref{FIG1}].
In analogy with the conventional BPE,
the currents arising from the diagonal and off-diagonal components of $\rho_{mn}^{(2;EB)}$ are termed the injection-like and shift-like currents, respectively,
each likewise containing $\mathcal{T}$-even and $\mathcal{T}$-odd contributions [lower panel of Fig.~\ref{FIG1}].

For the diagonal component $\rho_{mm}^{(2;EB)}$, we define
$\partial_t j^a_{\rm inj}
=
2\kappa^{abc}_{\rm inj;+}
{\rm Re}
\left[
E_b^\omega\bar{B}_{c}^{\omega*}
\right]
+
2\kappa^{abc}_{\rm inj;-}
{\rm Im}
\left[
E_b^{\omega}\bar{B}_{c}^{\omega*}
\right]$,
where $E_b^\omega=E_b(\omega)$ and $\bar{B}_c^{\omega*}=\bar{B}_c(-\omega)$.
The $\mathcal{T}$-even and $\mathcal{T}$-odd MBPE injection-like susceptibilities are then given by~\cite{sup}
\begin{align}
\kappa^{abc}_{\rm inj;+}
&=
\frac{2\pi e^2}{\hbar^2}
\sum_{nm}
\int_{k}
f_{mn}\Delta^a_{nm}
\mathcal{Q}^{bc}_{nm}
\delta(\hbar\omega-\epsilon_{mn}),
\label{eq:eta_plus}
\\
\kappa^{abc}_{\rm inj;-}
&=
\frac{\pi e^2}{\hbar^2}
\sum_{nm}
\int_{k}
f_{mn}\Delta^a_{nm}
\mathcal{Z}^{bc}_{mn}
\delta(\hbar\omega-\epsilon_{mn}),
\label{eq:eta_minus}
\end{align}
where $e$ and $\hbar$ are restored by dimensional analysis,
$\epsilon_{mn}=\epsilon_{m}-\epsilon_n=-\epsilon_{nm}$,
$\Delta^a_{nm}=\partial_a\epsilon_{nm}$, with $\partial_a=\partial/\partial k_a$,
and $f_{nm}=f_n-f_m$, with $f_n$ the equilibrium Fermi distribution function.
In addition, $\mathcal{Q}^{bc}_{nm}=\mathrm{Re}[r^b_{nm}\sigma^c_{mn}]$ and $\mathcal{Z}^{bc}_{nm}=-2\mathrm{Im}[r^b_{nm}\sigma^c_{mn}]$
are the Zeeman quantum metric and Zeeman Berry curvature~\cite{xiangZeeman}, respectively.
Here, $r^b_{nm}=\langle n|i\partial_b|m\rangle$ with $n \neq m$ is the interband Berry connection,
while $\sigma^c_{mn}=\langle m|\hat{\sigma}^c|n\rangle$, with $\hat{\sigma}^c$ the Pauli matrix in spin space.
The $\mathcal{T}$-even $\kappa^{abc}_{\mathrm{inj};+}$, governed by the Zeeman quantum metric dipole (ZQMD) $\Delta^a_{nm}\mathcal{Q}^{bc}_{nm}$,
shares the same quantum-geometric origin as the spin-induced magnetononlinear Hall effect~\cite{xiangMap,Xiang2024IMHE}
and the extrinsic nonlinear spin Edelstein effect~\cite{xiangSpin}.
In contrast, the $\mathcal{T}$-odd $\kappa^{abc}_{\mathrm{inj};-}$, governed by the Zeeman Berry curvature dipole (ZBCD) $\Delta^a_{nm}\mathcal{Z}^{bc}_{nm}$,
shares the same quantum-geometric origin as the spin-induced planar Hall effect~\cite{xiangMap} and the intrinsic nonlinear spin Edelstein effect~\cite{xiangSpin}.

For the off-diagonal component $\rho_{mn}^{(2;EB)}$ with $m \neq n$,
we define
$j^a_{\rm sht} = 2\kappa^{abc}_{\rm sht;-} {\rm Re} \left[ E_b^{\omega}\bar{B}_{c}^{\omega*} \right]
+
2\kappa^{abc}_{\rm sht;+} {\rm Im} \left[ E_b^{\omega}\bar{B}_{c}^{\omega*} \right]$.
The $\mathcal{T}$-even and $\mathcal{T}$-odd MBPE shift-like susceptibilities are then given by~\cite{sup}
\begin{align}
\kappa^{abc}_{\rm sht;+}
&=
+ \frac{\pi e^2}{\hbar}
\sum_{nm}
\int_{k}
f_{nm}
\text{Re}\left[ \mathcal{K}^{abc}_{mn} \right]
\delta(\hbar\omega-\epsilon_{mn}),
\label{eq:zeta_plus}
\\
\kappa^{abc}_{\rm sht;-}
&=
-\frac{\pi e^2}{\hbar}
\sum_{nm}
\int_{k}
f_{nm}
\text{Im}\left[\mathcal{K}^{abc}_{mn}\right]
\delta(\hbar\omega-\epsilon_{mn}).
\label{eq:zeta_minus}
\end{align}
Here, $\mathcal{K}^{abc}_{mn}\equiv \sigma^c_{mn}r^b_{nm;a}-\sigma^c_{mn;a}r^b_{nm}$,
where $r^b_{nm;a}=\mathcal{D}^a_{nm}r^b_{nm}$ and $\sigma^c_{mn;a}=\mathcal{D}^a_{mn}\sigma^c_{mn}$.
The $U(1)$ covariant derivative is defined as $\mathcal{D}^a_{mn}=\partial_a-i(\mathcal{A}^a_{mm}-\mathcal{A}^a_{nn})$,
with $\mathcal{A}^a_{mm}=\langle m|i\partial_a|m\rangle$ the intraband Berry connection.
Using the sum rules for $r^b_{nm;a}$~\cite{SumRule} and $\sigma^c_{mn;a}$~\cite{xiangSpin}, and neglecting three-band processes~\cite{Zhang2019},
we find $\mathcal{K}^{abc}_{mn}=\sigma^c_{mn}\left(\Delta^a_{nm}r^b_{nm}+r^a_{nm}\Delta^b_{nm}+iw^{ab}_{nm}\right)/\epsilon_{nm}$,
where $w^{ab}_{nm}=\langle n|\partial^2_{ab}\hat{H}_0|m\rangle$ typically gives a subleading contribution~\cite{xiangThird2026}.
Consequently, the $\mathcal{T}$-even $\kappa^{abc}_{\mathrm{sht};+}$ also receives contributions from the ZQMD,
$\Delta^a_{nm}\mathcal{Q}^{bc}_{nm}+\Delta^b_{nm}\mathcal{Q}^{ac}_{nm}$,
whereas the $\mathcal{T}$-odd $\kappa^{abc}_{\mathrm{sht};-}$ receives contributions from the ZBCD,
$\Delta^a_{nm}\mathcal{Z}^{bc}_{nm}+\Delta^b_{nm}\mathcal{Z}^{ac}_{nm}$.

Eqs.~\eqref{eq:eta_plus}--\eqref{eq:zeta_minus} are the main results of this Letter.
Two remarks are in order.
First, beyond their distinct parities under $\mathcal{P}$,
the BPE and MBPE exhibit complementary relations between light polarization and $\mathcal{T}$ symmetry,
as summarized in Fig.~\ref{FIG1}.
Under linearly polarized light (LPL), the conventional injection and shift susceptibilities are $\mathcal{T}$-odd and $\mathcal{T}$-even, respectively,
whereas the corresponding MBPE injection-like and shift-like susceptibilities are $\mathcal{T}$-even and $\mathcal{T}$-odd.
Under circularly polarized light (CPL), the time-reversal parities of both susceptibilities are reversed.
Second, all four MBPE susceptibilities are gauge invariant
and can be evaluated for realistic materials using state-of-the-art first-principles methods~\cite{Pizzi2020,Xiao2021,WannierNLQG},
facilitating the experimental search for the MBPE.

\begin{figure}[t!]
\includegraphics[width=0.95\columnwidth]{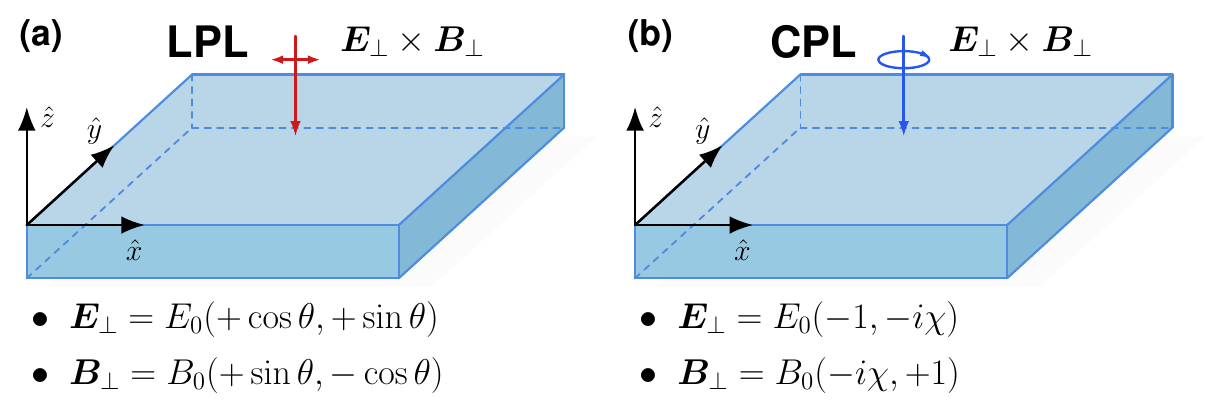}
\caption{Illumination configurations for a two-dimensional system under normal incidence with (a) LPL and (b) CPL.}
\label{FIG2}
\end{figure}

\bigskip
\noindent{\textit{\textcolor{blue}{Illumination configuration and symmetry analysis.}}}---In addition to the spin Zeeman coupling,
the magnetic component of light can also couple to the orbital motion through $\bm{B}\cdot(\bm{r}\times\bm{v})$.
To focus on the spin-Zeeman contribution,
we consider a two-dimensional system in the $xy$ plane under normally incident light propagating along $-\hat{z}$,
as illustrated in Fig.~\ref{FIG2}.
In this configuration, both $\bm{E}_{\perp}$ and $\bm{B}_{\perp}$ lie in the $xy$ plane,
whereas $(\bm{r}\times\bm{v})\parallel\hat{z}$.
Consequently, $\bm{B}_{\perp}\cdot(\bm{r}\times\bm{v})=0$, suppressing the orbital contribution.
Since $\vect{E}_{\perp}$, $\vect{B}_{\perp}$, and $\vect{j}$ are all in-plane,
we need only consider the MBPE components with $a,b,c\in\{x,y\}$, namely
\begin{equation}
[\kappa_{\pm}^{abc}]\equiv
\begin{pmatrix}
\kappa^{xxx}_{\pm} & \kappa^{xxy}_{\pm} & \kappa^{xyx}_{\pm} & \kappa^{xyy}_{\pm} \\
\kappa^{yxx}_{\pm} & \kappa^{yxy}_{\pm} & \kappa^{yyx}_{\pm} & \kappa^{yyy}_{\pm}
\end{pmatrix},
\label{eq:tensor_shape}
\end{equation}
where $\kappa_{\pm}^{abc} \in \{\kappa_{\mathrm{inj};\pm}^{abc}, \kappa_{\mathrm{sht};\pm}^{abc}\}$.
The components in Eq.~\eqref{eq:tensor_shape} are further constrained by magnetic point-group symmetries.
In particular, under a unitary symmetry $R$ or an antiunitary symmetry $R\mathcal{T}$,
the $\mathcal{T}$-even susceptibilities $\kappa_{+}^{abc}$ transform as
$\kappa^{abc}_{+} = \det(R) R_{aa'}R_{bb'}R_{cc'} \kappa^{a'b'c'}_{+}$,
whereas the $\mathcal{T}$-odd susceptibilities transform as $\kappa_{-}^{abc} = \eta_{\mathcal{T}} \det(R) R_{aa'}R_{bb'}R_{cc'} \kappa_{-}^{a'b'c'}$,
where $\eta_{\mathcal{T}}=+1$ for $R$ and $-1$ for $R\mathcal{T}$.
A useful constraint follows immediately from $R=C_{2z}$.
For $a,b,c\in\{x,y\}$,
$C_{2z} \kappa^{abc}_{\pm} \rightarrow -\kappa^{abc}_{\pm}$,
so that an unprimed $C_{2z}$ forbids all in-plane MBPE components.
For $C_{2z}\mathcal{T}$, $\kappa^{abc}_{+}$ remains forbidden,
whereas $\kappa^{abc}_{-}$ can survive since $C_{2z}\mathcal{T}\kappa^{abc}_{-} \rightarrow \kappa^{abc}_{-}$.

Furthermore, because $\kappa_{\pm}^{abc}$ are even under $\mathcal{P}$,
their symmetry constraints can be classified according to the 32 magnetic Laue groups (MLGs),
comprising 11 type-I, 11 type-II, and 10 type-III groups,
as detailed in the Supplemental Material~\cite{sup}.
For the $\mathcal{T}$-even MBPE in time-reversal-invariant systems,
it suffices to consider the 11 type-II MLGs, which contain $\mathcal{T}$ as an independent symmetry.
With the principal rotation axis chosen along $\hat{z}$,
eight of these MLGs contain an unprimed $C_{2z}$,
either explicitly or generated as a power or product of the group generators,
and hence forbid all eight in-plane components in Eq.~\eqref{eq:tensor_shape}.
The allowed type-II MLGs are $\bar{1}1'$, $\bar{3}1'$, and $\bar{3}m1'$.
For the $\mathcal{T}$-even susceptibilities $\kappa_{+}^{abc}$,
with Jahn notation $eV^3$,
the Bilbao Crystallographic Server~\cite{Bilbao} gives eight, two,
and one independent components for $\bar{1}1'$, $\bar{3}1'$, and $\bar{3}m1'$, respectively,
as summarized in TABLE~\ref{tab:MLG_MBPE}.

The $\mathcal{T}$-odd susceptibilities $\kappa_{-}^{abc}$ vanish identically in all type-II MLGs because $\mathcal{T}$ is an independent symmetry.
Among the 11 type-I MLGs, only $\bar{1}$, $\bar{3}$, and $\bar{3}m$ allow purely in-plane components of $\kappa_{-}^{abc}$,
with the same spatial constraints as $\bar{1}1'$, $\bar{3}1'$, and $\bar{3}m1'$ for $\kappa_{+}^{abc}$, respectively.
Among the ten type-III MLGs, six contain an unprimed $C_{2z}$ and are therefore excluded.
The remaining four are $2'/m'$, $\bar{3}m'$, $6'/m'$, and $6'/m'mm'$.
Using the Jahn notation $aeV^3$ for the $\mathcal{T}$-odd susceptibilities $\kappa_{-}^{abc}$,
we obtain the corresponding allowed components from the Bilbao Crystallographic Server~\cite{Bilbao}
and summarize them in TABLE~\ref{tab:MLG_MBPE}.
For completeness, TABLE~\ref{tab:MLG_MBPE} also lists the allowed components of $\kappa_{-}^{abc}$ for $\bar{1}$, $\bar{3}$, and $\bar{3}m$.

To close this section, we wish to remark that both $\mathcal{T}$-even and $\mathcal{T}$-odd MBPEs can be accessed with appropriate light polarization.
However, for the CPL configuration considered here [Fig.~\ref{FIG2}(b)],
we have $\operatorname{Im}\!\left[E_x^\omega \bar{B}_y^{\omega *}\right] =\operatorname{Im}\!\left[E_y^\omega \bar{B}_x^{\omega *}\right]=0$,
whereas $\operatorname{Im}\!\left[E_x^\omega \bar{B}_x^{\omega *}\right] =\operatorname{Im}\!\left[E_y^\omega \bar{B}_y^{\omega *}\right]\neq0$.
Consequently, the CPL-induced MBPE current is $j^a = 2(\kappa_{\pm}^{axx}+\kappa_{\pm}^{ayy}) \operatorname{Im}\!\left[E_x^\omega \bar{B}_x^{\omega *}\right]$,
which vanishes when $\kappa_{\pm}^{axx}=-\kappa_{\pm}^{ayy}$,
as occurs for all trigonal and hexagonal MLGs listed in TABLE~\ref{tab:MLG_MBPE}.
Representative two-dimensional materials in these symmetry classes include
nonmagnetic monolayers silicene~\cite{Liu2011PRL,Liu2011PRB},
$\mathrm{1T}$-PtX$_2$ (X=S, Se, Te)~\cite{PtSe2_exp,PtTe2_exp,PtS2_exp},
and jacutingaite Pt$_2$HgSe$_3$~\cite{Marrazzo2018},
as well as ferromagnetic monolayers CrI$_3$~\cite{CrI31,CrI32,CrI3_exp},
Cr$_2$Ge$_2$Te$_6$~\cite{Cr2Ge2Te6_exp}, $\mathrm{1T}$-FeCl$_2$~\cite{FeCl2_exp}, and CrSiTe$_3$~\cite{Lin2016,Fujita2022,Kang2023}.
This cancellation, namely $\kappa_{\pm}^{axx}+\kappa_{\pm}^{ayy}=0$,
however, is not generic to lower-symmetry magnetic systems.
In particular, the $2'/m'$ MLG permits a finite CPL-induced MBPE.
This symmetry is relevant to several predicted two-dimensional $d$-wave altermagnets,
including RuF$_4$, VF$_4$, and AgF$_2$~\cite{Sodequist2024}.
Accordingly, for the $\mathcal{T}$-even MBPE we focus on the LPL-induced injection-like susceptibility $\kappa_{\rm inj;+}^{abc}$,
whereas for the $\mathcal{T}$-odd MBPE we employ a RuF$_4$-motivated altermagnetic model to
examine the LPL-induced shift-like and CPL-induced injection-like MBPEs.

\begin{table}[t]
\caption{Symmetry-allowed in-plane MBPE susceptibilities $\kappa_{+}^{abc}$ and $\kappa_{-}^{abc}$, with $a,b,c\in\{x,y\}$.
Here $\kappa_{+}^{abc}$ denote the $\mathcal{T}$-even MBPE susceptibilities in $\mathcal{T}$-invariant systems
and are classified by the type-II MLGs using the Jahn notation $eV^3$,
whereas $\kappa_{-}^{abc}$ denote the $\mathcal{T}$-odd MBPE susceptibilities in $\mathcal{T}$-breaking systems
and are classified by the type-I and type-III MLGs using $aeV^3$. Only MLGs allowing nonzero components are listed.
Here the principal rotation axis is chosen along $\hat{z}$.
\label{tab:MLG_MBPE}}
\centering
\scriptsize
\renewcommand{\arraystretch}{1.18}
\setlength{\tabcolsep}{2.5pt}

\begin{tabular}{c|c|l}
\hline\hline

MBPE & MLGs & Tensor form \\
\hline

\multirow{3}{*}{$\kappa_{+}^{abc}\;(eV^3)$}
&
$\bar{1}1'$
&
All 8 independent components
\\
\cline{3-3}

&
$\bar{3}1'$
&
$\begin{gathered}
xxx=-xyy=-yxy=-yyx,\\[-1mm]
yyy=-yxx=-xyx=-xxy
\end{gathered}$
\\
\cline{3-3}

&
$\bar{3}m1'$
&
$xxx=-xyy=-yxy=-yyx$
\\

\hline

\multirow{4}{*}{$\kappa_{-}^{abc}\;(aeV^3)$}
&
$\bar{1},\,2'/m'$
&
All 8 independent components
\\
\cline{3-3}

&
$\bar{3},\,6'/m'$
&
$\begin{gathered}
xxx=-xyy=-yxy=-yyx,\\[-1mm]
yyy=-yxx=-xyx=-xxy
\end{gathered}$
\\
\cline{3-3}

&
$\bar{3}m,\,6'/m'mm'$
&
$xxx=-xyy=-yxy=-yyx$
\\
\cline{3-3}

&
$\bar{3}m'$
&
$yyy=-yxx=-xyx=-xxy$
\\

\hline\hline
\end{tabular}
\end{table}

\begin{figure}[t!]
\includegraphics[width=0.90\columnwidth]{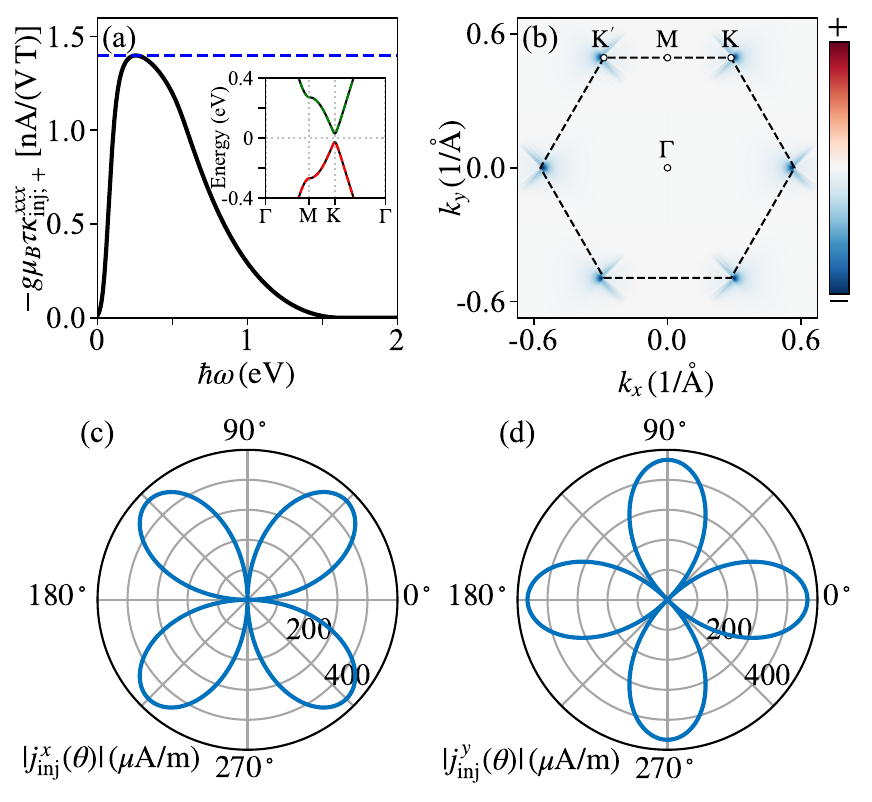}
\caption{$\mathcal{T}$-even injection-like MBPE in monolayer jacutingaite Pt$_2$HgSe$_3$.
(a) Frequency dependence of the effective injection-like MBPE susceptibility $-g\mu_B\tau\kappa_{\rm inj;+}^{xxx}$.
The inset shows the corresponding band structure along the high-symmetry path $\Gamma$--$\mathrm{M}$--$\mathrm{K}$--$\Gamma$.
The blue dashed line marks $1.4~{\rm nA/(V\,T)}$, the value used to estimate the currents in panels (c) and (d).
(b) Momentum-space distribution of the ZQMD $\Delta_{cv}^{x}\mathcal{Q}_{cv}^{xx}$.
The dashed hexagon denotes the first Brillouin zone and the high-symmetry points $\Gamma$, $\mathrm{K}$, $\mathrm{K}'$, and $\mathrm{M}$ are indicated.
(c,d) Angular dependences of $|j^x(\theta)|$ and $|j^y(\theta)|$, respectively,
obtained from Eqs.~\eqref{jx} and~\eqref{jy} using $E_0=10^7~{\rm V/m}$ and $B_0=E_0/c$ with $c=3\times10^8~{\rm m/s}$.
The corresponding current-density amplitude is approximately $467~\mu{\rm A/m}$.
}
\label{FIG3}
\end{figure}

\bigskip
\noindent{\textcolor{blue}{\textit{MBPE in monolayer jacutingaite Pt$_2$HgSe$_3$.}}}---We first illustrate the $\mathcal{T}$-even MBPE
in monolayer jacutingaite Pt$_2$HgSe$_3$~\cite{Marrazzo2018}, a large-gap and switchable Kane--Mele quantum spin Hall insulator.
Its low-energy electronic structure is described by a buckled Kane--Mele model~\cite{Marrazzo2020},
\begin{align}
\hat{H}_0=
&-t\sum_{\langle ij\rangle\sigma}
c_{i\sigma}^{\dagger}c_{j\sigma}
+i\Delta\sum_{\langle\!\langle ij\rangle\!\rangle\sigma\sigma'}
\nu_{ij}
c_{i\sigma}^{\dagger}
(s_z)_{\sigma\sigma'}
c_{j\sigma'}
\nonumber\\
&+i\Delta'\sum_{\langle\!\langle ij\rangle\!\rangle\sigma\sigma'}
u_{ij}
c_{i\sigma}^{\dagger}
\bigl(\bm{s}\times\bm{d}_{ij}^{0}\bigr)_z{}_{\sigma\sigma'}
c_{j\sigma'}.
\label{eq:KM}
\end{align}
Here $t$ is the nearest-neighbor hopping,
while $\Delta$ and $\Delta'$ denote the intrinsic and in-plane second-neighbor spin-orbit couplings (SOCs), respectively.
The factor $\nu_{ij}=[(\bm d_1\times\bm d_2)/|\bm d_1\times\bm d_2|]_z=\pm1$
specifies the orientation of the two-step nearest-neighbor path connecting the second-neighbor sites $i$ and $j$,
where $\bm d_1$ and $\bm d_2$ are the two successive nearest-neighbor bond vectors along the path,
and $u_{ij}=\pm1$ distinguishes the two sublattices.
The unit vector $\bm d_{ij}^{0}=(\bm r_j-\bm r_i)/|\bm r_j-\bm r_i|$ points from site $i$ to site $j$.
For numerical calculations,
we adopt the representative parameters $t=0.27$ eV and $\Delta/t=\Delta'/t=0.02$ used in Ref.~\onlinecite{Marrazzo2020},
corresponding to $\Delta=\Delta'=5.4$ meV,
and use a Gaussian broadening of $30$~meV for the resonant delta function.

Monolayer jacutingaite Pt$_2$HgSe$_3$ has the crystallographic point group $D_{3d}=\bar{3}m$~\cite{Marrazzo2020}.
In contrast to a planar honeycomb lattice,
the absence of horizontal mirror symmetry allows the in-plane SOC term proportional to $\Delta'$,
while both spatial inversion $\mathcal{P}$ and time-reversal symmetry $\mathcal{T}$ remain preserved.
Accordingly, the corresponding type-II magnetic Laue group is $\bar{3}m1'$.
As shown in TABLE~\ref{tab:MLG_MBPE},
this symmetry allows only one independent in-plane $\mathcal{T}$-even MBPE susceptibility component,
$\kappa_{+}^{xxx} =-\kappa_{+}^{xyy} =-\kappa_{+}^{yxy} =-\kappa_{+}^{yyx}$,
with all other in-plane components vanishing.
This constraint applies separately to both $\kappa_{\rm inj;+}^{abc}$ and $\kappa_{\rm sht;+}^{abc}$.
For the normally incident CPL configuration, however, the $\mathcal{T}$-even shift-like response vanishes,
as discussed above.
We therefore focus on the LPL-induced injection-like MBPE, for which we find~\cite{footnote1}
\begin{align}
j^x_{\rm inj} 
&=
-g \mu_B \tau \kappa_{\rm inj;+}^{xxx} E_0B_0\sin2\theta, \label{jx}
\\
j^y_{\rm inj} 
&= 
-g \mu_B \tau \kappa_{\rm inj;+}^{xxx} E_0B_0\cos2\theta, \label{jy}
\end{align}
where $\tau=10^{-14}\,\mathrm{s}$ is adopted,
within the experimentally relevant $10^{-14}$--$10^{-12}\,\mathrm{s}$ range for photoexcited carrier relaxation~\cite{Burger2019}.
Consequently, it suffices to evaluate the single independent component $\kappa_{\rm inj; +}^{xxx}$
to characterize the in-plane $\mathcal{T}$-even MBPE of this model.

Figure~\ref{FIG3}(a) shows the frequency dependence of the effective injection-like MBPE susceptibility $-g\mu_B\tau\kappa_{\rm inj;+}^{xxx}$.
It turns on above the direct-gap absorption threshold at the $\mathrm{K}$ point [see the inset of Fig.~\ref{FIG3}(a)],
reaches a maximum of about $1.4~{\rm nA/(V\,T)}$ at $\hbar\omega\simeq0.25$ eV,
and then gradually decreases with increasing photon energy.
The corresponding ZQMD $\Delta_{cv}^{x}\mathcal{Q}_{cv}^{xx}$ between the conduction and valence doublets is strongly concentrated near the
$\mathrm{K}$ and $\mathrm{K}'$ valleys, as shown in Fig.~\ref{FIG3}(b),
indicating that the response is dominated by valley-localized contributions from the Zeeman quantum geometry~\cite{xiangZeeman, xiangSpin}.
Using $-g\mu_B\tau\kappa_{\rm inj;+}^{xxx}=1.4~{\rm nA/(V\,T)}$, $E_0=10^7~{\rm V/m}$, and $B_0=E_0/c$ with $c=3\times10^8~{\rm m/s}$,
Eqs.~\eqref{jx} and~\eqref{jy} give a characteristic current-density amplitude of about $467~\mu{\rm A/m}$.
For an illuminated width of $10~\mu{\rm m}$, this corresponds to a photocurrent of approximately $4.7~\rm{nA}$,
comparable to experimentally measurable photocurrents in the conventional BPE~\cite{Li2021CIPS}.

Figures~\ref{FIG3}(c) and~\ref{FIG3}(d) further show the polarization dependence of the MBPE.
The magnitude $|j^{x}_{\rm inj}|$ reaches its maxima at $\theta=45^\circ,135^\circ,\cdots$,
whereas $|j^{y}_{\rm inj}|$ is maximal at $\theta=0^\circ,90^\circ,\cdots$.
The resulting four-lobe patterns are therefore rotated by $45^\circ$ with respect to each other.
Such polarization anisotropy can be directly measured to characterize the photovoltaic response and extract the underlying susceptibility~\cite{Liang2023}.
In the present case, it provides a direct means to determine the single independent component $\kappa_{\rm inj;+}^{xxx}$.
Notably, this polarization-dependent photocurrent arises in a centrosymmetric solid,
where the conventional electric-dipole BPE is forbidden by inversion symmetry.

\bigskip
\noindent \textit{\textcolor{blue}{Discussion and summary.}}---To further assess the material applicability of the $\mathcal{T}$-even MBPE,
we have also performed calculations for the first-principles tight-binding model of monolayer PtSe$_2$~\cite{xiangPtSe2}.
As detailed in the Supplemental Material~\cite{sup},
the resulting $\mathcal{T}$-even injection-like MBPE reaches a current density about one order of magnitude larger than that obtained for monolayer Pt$_2$HgSe$_3$.
Sizable $\mathcal{T}$-odd MBPEs can likewise arise in magnetic materials.
As discussed in the End Matter using an effective tight-binding model of altermagnetic monolayer RuF$_4$,
both the $\mathcal{T}$-odd injection-like and shift-like responses reach several ${\rm nA/(V\,T)}$,
corresponding to current densities on the order of $1~{\rm mA/m}$ under experimentally accessible optical fields,
larger than the $\mathcal{T}$-even MBPE discussed above.
Notably, the CPL-induced injection-like current reverses sign upon switching either the light helicity or the N\'eel vector,
whereas the LPL-induced shift-like current exhibits a characteristic polarization dependence and reverses upon reversal of the N\'eel vector.

The MBPE is not restricted to centrosymmetric solids.
In noncentrosymmetric solids, the conventional BPE and MBPE are both symmetry allowed.
In particular, as shown in Fig.~\ref{FIG1},
the $\mathcal{T}$-even injection-like current and the conventional $\mathcal{T}$-even shift current can coexist under LPL.
Exploring the interplay between these two distinct photovoltaic mechanisms is an interesting topic for future studies.
In addition, although we have focused on the spin-Zeeman contribution to the MBPE,
exploring its orbital contribution is an interesting direction for future work.
Beyond the charge photocurrent considered here, the present framework can also be extended to the spin degree of freedom,
in analogy with the bulk spin photovoltaic effect~\cite{Xu2021BSPV},
providing a possible route to a magnetoelectric spin bulk photovoltaic effect in centrosymmetric solids.

In summary, we establish the magnetoelectric bulk photovoltaic effect,
a dc photocurrent generated by the bilinear coupling between the electric and magnetic components of light.
In contrast to the conventional electric-dipole BPE,
the MBPE remains allowed in centrosymmetric solids.
We identify four injection-like and shift-like MBPE responses distinguished by time-reversal symmetry and light polarization,
and reveal their microscopic origins in the dipole moments of the Zeeman quantum metric and Zeeman Berry curvature.
Our symmetry analysis and material calculations demonstrate that sizable MBPEs can arise in broad classes of nonmagnetic and magnetic materials
with experimentally accessible current densities.
These results extend the bulk photovoltaic effect beyond the conventional requirement of broken inversion symmetry
and open a route to probing Zeeman quantum geometry and altermagnetic order through photocurrent responses.

\bigskip
\noindent \textit{\textcolor{blue}{Acknowledgements.}}---We thank the National Natural Science Foundation of China (Grants Nos. 12674063, 12404059, 12574054, and 12474047).

\newpage
\section*{End Matter}
\begin{figure}[t!]
\includegraphics[width=0.90\columnwidth]{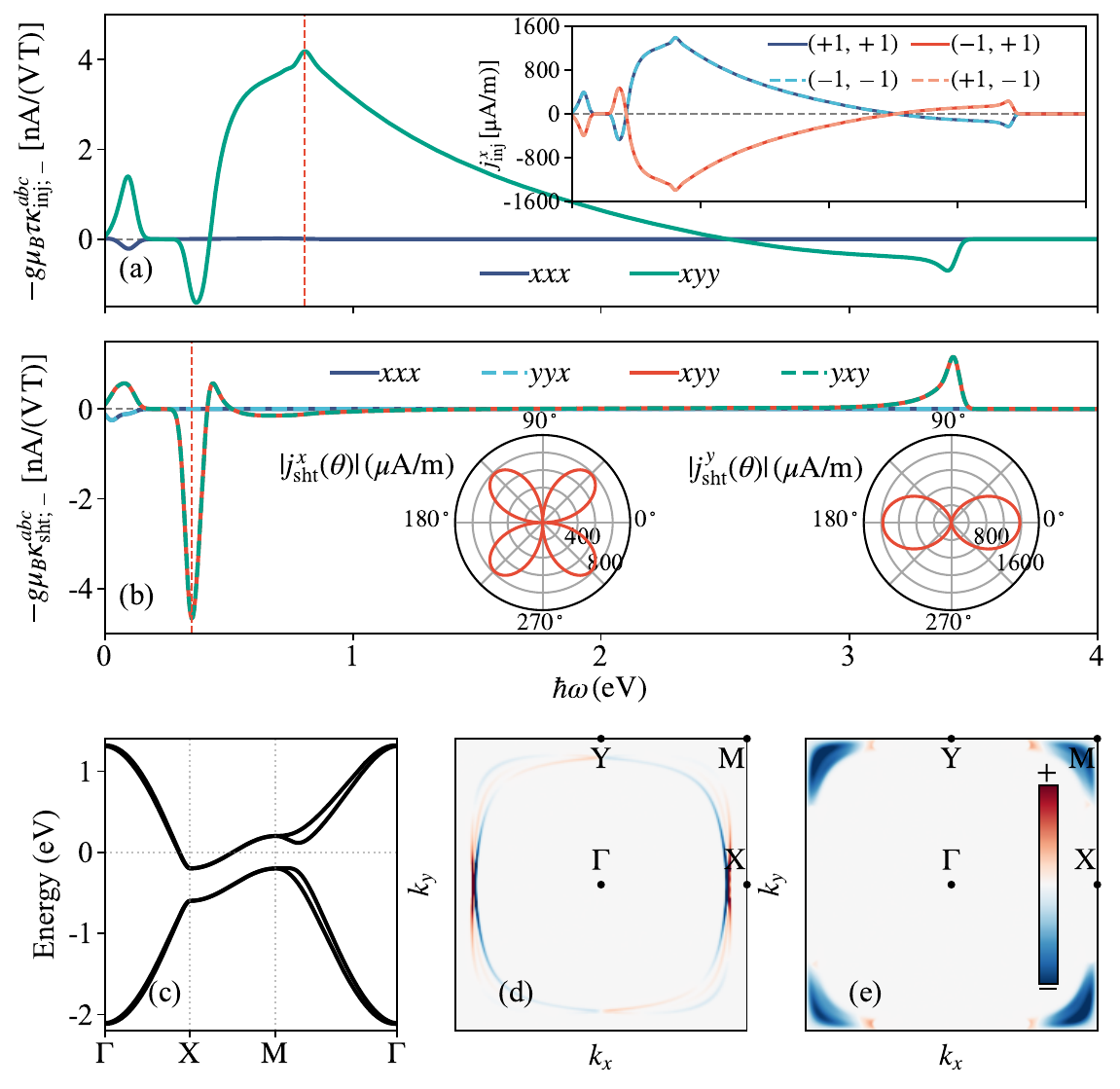}
\caption{$\mathcal{T}$-odd injection-like and shift-like MBPEs in monolayer RuF$_4$.
(a) Frequency dependence of the effective injection-like susceptibilities
$-g\mu_B\tau\kappa_{\rm inj;-}^{xxx}$ and
$-g\mu_B\tau\kappa_{\rm inj;-}^{xyy}$.
(b) Frequency dependence of the symmetry-allowed effective shift-like
susceptibilities
$-g\mu_B\kappa_{\rm sht;-}^{xxx}$,
$-g\mu_B\kappa_{\rm sht;-}^{yyx}$,
$-g\mu_B\kappa_{\rm sht;-}^{xyy}$, and
$-g\mu_B\kappa_{\rm sht;-}^{yxy}$.
The vertical dashed lines in (a) and (b) indicate
$\hbar\omega=0.806$~eV and $0.350$~eV, respectively.
(c) Band structure along the high-symmetry path
$\Gamma$--$\mathrm{X}$--$\mathrm{M}$--$\Gamma$.
(d) Momentum-resolved integrand of
$-g\mu_B\tau\kappa_{\rm inj;-}^{xyy}$
at $\hbar\omega=0.806$~eV.
(e) Momentum-resolved integrand of
$-g\mu_B\kappa_{\rm sht;-}^{xyy}$
at $\hbar\omega=0.350$~eV.
Red and blue indicate positive and negative contributions, respectively.
}
\label{FIG4}
\end{figure}

\noindent{\textit{\textcolor{blue}{MBPE in monolayer RuF$_4$.}}}---We illustrate the $\mathcal{T}$-odd MBPE using a model motivated by monolayer RuF$_4$,
a predicted two-dimensional $d$-wave altermagnet with the magnetic space group $P2_1'/c'$~\cite{Sodequist2024,Peng2026}.
We model its low-energy electronic structure using the symmetry-based minimal two-sublattice Hamiltonian~\cite{Gondolf2025}
\begin{equation}
\hat{H}_0
=
\varepsilon_0\hat{\tau}_0\hat{\sigma}_0
+
t_x^R\hat{\tau}_x\hat{\sigma}_0
-
t_x^I\hat{\tau}_y\hat{\sigma}_0
+
t_z\hat{\tau}_z\hat{\sigma}_0
+
\hat{\tau}_z\vect{N}\cdot\hat{\vect{\sigma}},
\end{equation}
where $\hat{\tau}_i$ and $\hat{\sigma}_i$ are Pauli matrices acting on the sublattice and spin degrees of freedom, respectively,
and $\vect{N}$ denotes the N\'eel order.  The momentum-dependent coefficients are $t_x^R=\operatorname{Re}[t_x]$ and $t_x^I=\operatorname{Im}[t_x]$,
with $t_x=-4t_1e^{-i(k_x+k_y)/2}\cos(k_x/2)\cos(k_y/2)$, $\varepsilon_0=-2t_2(\cos k_x+\cos k_y) -4t_3\cos k_x\cos k_y-\mu$, and $t_z=4t_4\sin k_x\sin k_y$.
The term $t_z$ transforms as $d_{xy}$ and, together with the staggered exchange $\hat{\tau}_z\vect{N}\cdot\hat{\vect{\sigma}}$,
produces the characteristic momentum-dependent altermagnetic spin splitting along $\mathrm{M}$--$\Gamma$ [see Fig.~\ref{FIG4}(c)].
In addition, we retain the symmetry-allowed SOC term $\hat{H}_{\rm soc}=\lambda\cos(k_x/2)\cos(k_y/2)\hat{\tau}_y\hat{\sigma}_z$,
which is compatible with the centrosymmetric two-sublattice minimal model for the $P2_1/c$ structure~\cite{Roig2024}.
For numerical calculations, we adopt the representative parameters $t_1=0.425$~eV, $t_2=0.05$~eV, $t_3=-0.025$~eV, $t_4=-0.075$~eV, $\mu=0.30$~eV,
and $|\vect{N}|=0.20$~eV used in Ref.~\onlinecite{Gondolf2025},
and take $\lambda=0.05$~eV as a representative SOC strength.

For monolayer RuF$_4$, the primed twofold axis lies in the plane,
which we choose along $y$, corresponding to the $2_y'/m_y'$ setting.
This differs from the $2_z'/m_z'$ setting adopted in TABLE~\ref{tab:MLG_MBPE},
where the principal rotation axis is chosen along $\hat{z}$ and all eight purely in-plane $\mathcal{T}$-odd components are independent.
For $2_y'/m_y'$, only four purely in-plane components remain allowed,
namely $\kappa_-^{xxx}$, $\kappa_-^{xyy}$, $\kappa_-^{yxy}$, and $\kappa_-^{yyx}$, which are mutually independent.
As a result, $\kappa_{\rm inj;-}^{xxx}+\kappa_{\rm inj;-}^{xyy}$ is not constrained to vanish,
so that the CPL-induced injection-like response remains symmetry allowed.
In particular, we have
\begin{align}
j^x_{\rm inj}=g\mu_B \tau (\kappa_{\rm inj;-}^{xxx}+\kappa_{\rm inj;-}^{xyy})\chi E_0B_0,
\end{align}
where $\chi=\pm1$ denotes the light helicity.
In the numerical calculations below,
we choose the in-plane N\'eel vector along $\hat{x}$, set $T=10$~K and $\tau=10^{-14}\,\mathrm{s}$~\cite{Burger2019},
and use a Gaussian broadening of $30$~meV for the resonant delta function.

Figure~\ref{FIG4}(a) shows the frequency dependence of the two effective susceptibilities entering the CPL-induced $\mathcal{T}$-odd injection-like current.
The response is strongly dominated by $-g\mu_B\tau\kappa_{\rm inj;-}^{xyy}$,
whereas the $xxx$ component remains much smaller throughout the frequency range.
The $xyy$ response exhibits several sign changes at low photon energies,
followed by a rapid enhancement above about $0.4$~eV.
It reaches a pronounced maximum of approximately $4.2~{\rm nA/(V\,T)}$ at $\hbar\omega\simeq0.806$~eV,
then gradually decreases and eventually changes sign again at higher energies.
The strong anisotropy between the $xyy$ and $xxx$ components originates from the in-plane N\'eel order $\vect{N}\parallel\hat{x}$.
In this configuration, $\hat{\sigma}_x$ is predominantly longitudinal with respect to the exchange field,
whereas $\hat{\sigma}_y$ is transverse and acquires much stronger interband matrix elements in the presence of SOC.
Consequently, the Zeeman Berry-curvature component $\mathcal{Z}^{yy}_{mn}$ entering $\Delta_{nm}^{x}\mathcal{Z}_{mn}^{yy}$
is strongly enhanced compared with the corresponding $\mathcal{Z}^{xx}_{mn}$ contribution,
accounting for the dominance of $\kappa_{\rm inj;-}^{xyy}$.

The microscopic origin of the maximum at $\hbar\omega\simeq0.806$~eV can be further identified from Figs.~\ref{FIG4}(c) and~\ref{FIG4}(d).
At this photon energy, the resonant interband condition $\epsilon_{mn}=\hbar\omega$ selects a closed contour near the Brillouin-zone boundary.
The momentum-resolved integrand is strongly concentrated along this contour and exhibits alternating positive and negative contributions.
Their incomplete cancellation, with the positive contribution prevailing upon integration over the Brillouin zone,
produces the pronounced positive peak in Fig.~\ref{FIG4}(a).

An important property of this $\mathcal{T}$-odd injection-like current is its double switchability.
Because $\kappa_{\rm inj;-}^{abc}$ changes sign upon reversal of the N\'eel vector, whereas the CPL response is also odd in the light helicity $\chi$,
one has $j_{\rm inj}^{x}\propto \chi\,\zeta$,
where $\zeta=\pm1$ labels the two opposite N\'eel-vector orientations.
Consequently, reversing either the light helicity or the N\'eel vector reverses the photocurrent,
while reversing both simultaneously leaves the current unchanged.
This behavior is illustrated in the inset of Fig.~\ref{FIG4}(a),
where the pairs $(\chi,\zeta)=(+1,+1)$ and $(-1,-1)$ give the same current,
whereas $(-1,+1)$ and $(+1,-1)$ give the opposite one.

Figure~\ref{FIG4}(b) shows the $\mathcal{T}$-odd shift-like MBPE.
For the present minimal model, the accidental $k_x\leftrightarrow k_y$ symmetry yields the numerical relations
$\kappa_{\rm sht;-}^{xxx}=\kappa_{\rm sht;-}^{yyx}$ and $\kappa_{\rm sht;-}^{xyy}=\kappa_{\rm sht;-}^{yxy}$.
These equalities are model specific rather than symmetry enforced,
since the $2_y'/m_y'$ magnetic Laue group generally allows the four components independently.
The response is again strongly anisotropic,
with the $xyy$ and $yxy$ components much larger than the $xxx$ and $yyx$ components.
This difference has the same microscopic origin as that of the injection-like response.
In particular, the dominant $yxy$ component contains the transverse spin matrix element $\sigma_y$,
whereas the much weaker $yyx$ component involves the predominantly longitudinal $\sigma_x$ channel for $\vect{N}\parallel\hat{x}$.

Note that $-g\mu_B\kappa_{\rm sht;-}^{xyy}$ reaches a pronounced minimum of about $-4.7~{\rm nA/(V\,T)}$ at $\hbar\omega\simeq0.350$~eV.
Figures~\ref{FIG4}(c) and~\ref{FIG4}(e) reveal the origin of this strong resonance.
The corresponding photon energy matches the small direct interband separation near the $\mathrm{M}$ point,
and the momentum-resolved shift-like integrand is strongly localized near the four $\mathrm{M}$-equivalent corners of the Brillouin zone.
Negative contributions dominate in these regions,
giving rise to the large negative peak in Fig.~\ref{FIG4}(b).
Since $\kappa_{\rm sht;-}^{abc}$ is $\mathcal{T}$-odd,
the entire shift-like photocurrent also reverses sign upon switching the N\'eel vector between the two opposite magnetic domains (not shown).

Unlike the CPL-induced $\mathcal{T}$-odd injection-like current,
the $\mathcal{T}$-odd shift-like current is not helicity dependent but remains strongly dependent on the polarization direction of LPL.
For the LPL illumination configuration in Fig.~\ref{FIG2}(a), the two in-plane current components are
\begin{align}
j_{\rm sht}^{x} &= -g\mu_B ( \kappa_{\rm sht;-}^{xxx} - \kappa_{\rm sht;-}^{xyy}) E_0B_0 \sin\theta\cos\theta,
\\
j_{\rm sht}^{y} &= -g\mu_B ( \kappa_{\rm sht;-}^{yyx}\sin^2\theta - \kappa_{\rm sht;-}^{yxy}\cos^2\theta)
E_0B_0.
\end{align}
Near the dominant resonance at $\hbar\omega\simeq0.350$~eV,
where the $xyy$ and $yxy$ components are equal and dominate the response,
these reduce approximately to $|j_{\rm sht}^{x}|\propto |\sin2\theta|$ and $|j_{\rm sht}^{y}|\propto \cos^2\theta$.
Accordingly, the inset of Fig.~\ref{FIG4}(b) displays a characteristic four-lobe pattern for $|j_{\rm sht}^{x}|$ and a two-lobe pattern for $|j_{\rm sht}^{y}|$.
Thus, although the $\mathcal{T}$-odd shift-like MBPE is not helicity dependent,
its magnitude and current direction can still be controlled by the linear polarization of the incident light.

Finally, using the same optical field strength as in the main text,
$E_0=10^7~{\rm V/m}$ and $B_0=E_0/c$, we have $E_0B_0\simeq3.33\times10^5~{\rm V\,T/m}$.
At the injection-like resonance,
the dominant effective susceptibility $-g\mu_B\tau\kappa_{\rm inj;-}^{xyy}\simeq4.2~{\rm nA/(V\,T)}$
corresponds to a characteristic current density of approximately $1.40~{\rm mA/m}$.
For the shift-like response, $|-g\mu_B\kappa_{\rm sht;-}^{yxy}|\simeq4.7~{\rm nA/(V\,T)}$
gives maximum current densities of approximately $0.78~{\rm mA/m}$ for $|j_{\rm sht}^{x}|$ and $1.58~{\rm mA/m}$ for $|j_{\rm sht}^{y}|$.
Assuming an effective illuminated width of $10~\mu{\rm m}$,
these values correspond to photocurrents of approximately $14.0~{\rm nA}$, $7.8~{\rm nA}$, and $15.8~{\rm nA}$, respectively.
These values are larger than the $\mathcal{T}$-even MBPE estimate discussed in the main text,
further supporting the experimental observability of the MBPE in centrosymmetric magnetic materials.


\begin{thebibliography}{00}
\bibitem{Fridkin2001}
V. M. Fridkin,
Bulk photovoltaic effect in noncentrosymmetric crystals,
\href{https://doi.org/10.1134/1.1387133}{Crystallogr. Rep. \textbf{46}, 654 (2001)}.
\bibitem{Alexe2011}
M. Alexe and D. Hesse,
Tip-enhanced photovoltaic effects in bismuth ferrite,
\href{https://doi.org/10.1038/ncomms1261}{Nat. Commun. \textbf{2}, 256 (2011)}.
\bibitem{Bhatnagar2013}
A. Bhatnagar, A. Roy Chaudhuri, Y. H. Kim, D. Hesse, and M. Alexe,
Role of domain walls in the abnormal photovoltaic effect in BiFeO$_3$,
\href{https://doi.org/10.1038/ncomms3835}{Nat. Commun. \textbf{4}, 2835 (2013)}.
\bibitem{Spanier2016}
J. E. Spanier, V. M. Fridkin, A. M. Rappe, A. R. Akbashev, A. Polemi, Y. Qi, Z. Gu, S. M. Young,
C. J. Hawley, D. Imbrenda, G. Xiao, A. L. Bennett-Jackson, and C. L. Johnson,
Power conversion efficiency exceeding the Shockley--Queisser limit in a ferroelectric insulator,
\href{https://doi.org/10.1038/nphoton.2016.143}{Nat. Photonics \textbf{10}, 611 (2016)}.
\bibitem{Osterhoudt2019}
G. B. Osterhoudt, L. K. Diebel, M. J. Gray, X. Yang, J. Stanco, X. Huang, B. Shen, N. Ni, P. J. W. Moll, Y. Ran, and K. S. Burch,
Colossal mid-infrared bulk photovoltaic effect in a type-I Weyl semimetal,
\href{https://doi.org/10.1038/s41563-019-0297-4}{Nat. Mater. \textbf{18}, 471 (2019)}.
\bibitem{Ma2019}
J. Ma, Q. Gu, Y. Liu, J. Lai, P. Yu, X. Zhuo, Z. Liu, J.-H. Chen, J. Feng, and D. Sun,
Nonlinear photoresponse of type-II Weyl semimetals,
\href{https://doi.org/10.1038/s41563-019-0296-5}{Nat. Mater. \textbf{18}, 476 (2019)}.
\bibitem{Han2021}
S. Han, M. Li, Y. Liu, W. Guo, M.-C. Hong, Z. Sun, and J. Luo,
Tailoring of a visible-light-absorbing biaxial ferroelectric towards broadband self-driven photodetection,
\href{https://doi.org/10.1038/s41467-020-20530-4}{Nat. Commun. \textbf{12}, 284 (2021)}.
\bibitem{Morimoto2016}
T. Morimoto and N. Nagaosa,
Topological nature of nonlinear optical effects in solids,
\href{https://doi.org/10.1126/sciadv.1501524}{Sci. Adv. \textbf{2}, e1501524 (2016)}.
\bibitem{deJuan2017}
F. de Juan, A. G. Grushin, T. Morimoto, and J. E. Moore,
Quantized circular photogalvanic effect in Weyl semimetals,
\href{https://doi.org/10.1038/ncomms15995}{Nat. Commun. \textbf{8}, 15995 (2017)}.
\bibitem{Ahn2020}
J. Ahn, G.-Y. Guo, and N. Nagaosa,
Low-frequency divergence and quantum geometry of the bulk photovoltaic effect in topological semimetals,
\href{https://doi.org/10.1103/PhysRevX.10.041041}{Phys. Rev. X \textbf{10}, 041041 (2020)}.
\bibitem{Ahn2022}
J. Ahn, G.-Y. Guo, N. Nagaosa, and A. Vishwanath,
Riemannian geometry of resonant optical responses,
\href{https://doi.org/10.1038/s41567-021-01465-z}{Nat. Phys. \textbf{18}, 290 (2022)}.
\bibitem{Ma2021}
Q. Ma, A. G. Grushin, and K. S. Burch,
Topology and geometry under the nonlinear electromagnetic spotlight,
\href{https://doi.org/10.1038/s41563-021-00992-7}{Nat. Mater. \textbf{20}, 1601 (2021)}.
\bibitem{Ma2023}
Q. Ma, R. Krishna Kumar, S.-Y. Xu, F. H. L. Koppens, and J. C. W. Song,
Photocurrent as a multiphysics diagnostic of quantum materials,
\href{https://doi.org/10.1038/s42254-022-00551-2}{Nat. Rev. Phys. \textbf{5}, 170 (2023)}.
\bibitem{vonBaltz1981}
R. von Baltz and W. Kraut,
Theory of the bulk photovoltaic effect in pure crystals,
\href{https://doi.org/10.1103/PhysRevB.23.5590}{Phys. Rev. B \textbf{23}, 5590 (1981)}.
\bibitem{Aversa1995}
C. Aversa and J. E. Sipe,
Nonlinear optical susceptibilities of semiconductors: Results with a length-gauge analysis,
\href{https://doi.org/10.1103/PhysRevB.52.14636}{Phys. Rev. B \textbf{52}, 14636 (1995)}.
\bibitem{Sipe2000}
J. E. Sipe and A. I. Shkrebtii,
Second-order optical response in semiconductors,
\href{https://doi.org/10.1103/PhysRevB.61.5337}{Phys. Rev. B \textbf{61}, 5337 (2000)}.
\bibitem{Young2012}
S. M. Young and A. M. Rappe,
First principles calculation of the shift current photovoltaic effect in ferroelectrics,
\href{https://doi.org/10.1103/PhysRevLett.109.116601}{Phys. Rev. Lett. \textbf{109}, 116601 (2012)}.
\bibitem{Wang2019}
H. Wang and X. Qian,
Ferroicity-driven nonlinear photocurrent switching in time-reversal invariant ferroic materials,
\href{https://doi.org/10.1126/sciadv.aav9743}{Sci. Adv. \textbf{5}, eaav9743 (2019)}.
\bibitem{Dai2023}
Z. Dai and A. M. Rappe,
Recent progress in the theory of bulk photovoltaic effect,
\href{https://doi.org/10.1063/5.0101513}{Chem. Phys. Rev. \textbf{4}, 011303 (2023)}.
\bibitem{Shi2021}
L.-K. Shi, D. Zhang, K. Chang, and J. C. W. Song,
Geometric photon-drag effect and nonlinear shift current in centrosymmetric crystals,
\href{https://doi.org/10.1103/PhysRevLett.126.197402}{Phys. Rev. Lett. \textbf{126}, 197402 (2021)}.
\bibitem{Ikeda2023}
Y. Ikeda, S. Kitamura, and T. Morimoto,
Photocurrent induced by a bicircular light drive in centrosymmetric systems,
\href{https://doi.org/10.1103/PhysRevLett.131.096301}{Phys. Rev. Lett. \textbf{131}, 096301 (2023)}.
\bibitem{Sivianes2025}
J. Sivianes, P. Garcia-Goiricelaya, D. Hernang\'omez-P\'erez, and J. Iba\~nez-Azpiroz,
Surface-state engineering for generation of nonlinear charge and spin photocurrents,
\href{https://doi.org/10.1103/h8rp-rtn8}{Phys. Rev. Lett. \textbf{135}, 256201 (2025)}.
\bibitem{Zhang2019}
Y. Zhang, T. Holder, H. Ishizuka, F. de Juan, N. Nagaosa, C. Felser, and B. Yan,
Switchable magnetic bulk photovoltaic effect in the two-dimensional magnet CrI$_3$,
\href{https://doi.org/10.1038/s41467-019-11832-3}{Nat. Commun. \textbf{10}, 3783 (2019)}.
\bibitem{Wang2020}
H. Wang and X. Qian,
Electrically and magnetically switchable nonlinear photocurrent in $PT$-symmetric magnetic topological quantum materials,
\href{https://doi.org/10.1038/s41524-020-00462-9}{npj Comput. Mater. \textbf{6}, 199 (2020)}.
\bibitem{Dong2026LayerPV}
Y. Dong, S. Kitamura, Y. M. Itahashi, D. G. Chica, S. Toyoda \textit{et al.},
Layer photovoltaic effect in a two-dimensional antiferromagnet with parity--time symmetry,
\href{https://doi.org/10.1038/s41563-026-02593-8}{Nat. Mater. \textbf{25}, 1776 (2026)}.
\bibitem{Ivchenko1984}
E. L. Ivchenko, Y. B. Lyanda-Geller, G. E. Pikus, and R. Y. Rasulov,
Magnetically induced photogalvanic effect in semiconductors,
Sov. Phys. Semicond. \textbf{18}, 55 (1984).
\bibitem{Burger2020}
A. M. Burger, L. Gao, R. Agarwal, A. Aprelev, J. E. Spanier, A. M. Rappe, and V. M. Fridkin,
Shift photovoltaic current and magnetically induced bulk photocurrent in piezoelectric sillenite crystals,
\href{https://doi.org/10.1103/PhysRevB.102.081113}{Phys. Rev. B \textbf{102}, 081113(R) (2020)}.
\bibitem{Hornung2021}
D. Hornung and R. von Baltz,
Quantum kinetics of the magnetophotogalvanic effect,
\href{https://doi.org/10.1103/PhysRevB.103.195203}{Phys. Rev. B \textbf{103}, 195203 (2021)}.
\bibitem{EMbook}
J. D. Jackson,
\textit{Classical Electrodynamics}, 3rd ed.  (Wiley, New York, 1999).
\bibitem{xiangZeeman}
L. Xiang, J. Jia, F. Xu, Z. Qiao, and J. Wang,
Intrinsic Gyrotropic Magnetic Current from Zeeman Quantum Geometry,
\href{https://doi.org/10.1103/PhysRevLett.134.116301}{Phys. Rev. Lett. \textbf{134}, 116301 (2025)}.
\bibitem{xiangSpin}
L. Xiang, H. Jin, and J. Wang,
Spin transport revealed by the spin quantum geometry,
\href{https://doi.org/10.1103/14mp-263q}{Phys. Rev. Lett. \textbf{135}, 146303 (2025)}.
\bibitem{xiangMap}
L. Xiang, J. Jia, F. Xu, and J. Wang,
Quantum geometric map of magnetotransport,
\href{https://doi.org/10.1103/kd12-4ycj}{Phys. Rev. B \textbf{113}, L201406 (2026)}.
\bibitem{Marrazzo2018}
A.~Marrazzo, M.~Gibertini, D.~Campi, N.~Mounet, and N.~Marzari,
Prediction of a Large-Gap and Switchable Kane-Mele Quantum Spin Hall Insulator,
\href{https://doi.org/10.1103/PhysRevLett.120.117701}{Phys. Rev. Lett. \textbf{120}, 117701 (2018)}.
\bibitem{Sodequist2024}
J.~S{\o}dequist and T.~Olsen,
Two-dimensional altermagnets from high throughput computational screening: Symmetry requirements, chiral magnons, and spin-orbit effects,
\href{https://doi.org/10.1063/5.0198285}{Appl. Phys. Lett. \textbf{124}, 182409 (2024)}.
\bibitem{AM2}
L. \v{S}mejkal, J. Sinova, and T. Jungwirth,
Beyond Conventional Ferromagnetism and Antiferromagnetism: A Phase with Nonrelativistic Spin and Crystal Rotation Symmetry,
\href{https://doi.org/10.1103/PhysRevX.12.031042}{Phys. Rev. X \textbf{12}, 031042 (2022)}.
\bibitem{AM3}
L. \v{S}mejkal, J. Sinova, and T. Jungwirth,
Emerging Research Landscape of Altermagnetism,
\href{https://doi.org/10.1103/PhysRevX.12.040501}{Phys. Rev. X \textbf{12}, 040501 (2022)}.
\bibitem{AM4}
L. Bai, W. Feng, S. Liu, L. \v{S}mejkal, Y. Mokrousov, and Y. Yao,
Altermagnetism: Exploring New Frontiers in Magnetism and Spintronics,
\href{https://doi.org/10.1002/adfm.202409327}{Adv. Funct. Mater. \textbf{34}, 2409327 (2024)}.
\bibitem{sup}
Supplemental Material.
\bibitem{Xiang2024IMHE}
L.~Xiang and J.~Wang,
Intrinsic in-plane magnetononlinear Hall effect in tilted Weyl semimetals,
\href{https://doi.org/10.1103/PhysRevB.109.075419}{Phys. Rev. B \textbf{109}, 075419 (2024)}.
\bibitem{SumRule}
A. M. Cook, B. M. Fregoso, F. de Juan, S. Coh, and J. E. Moore,
Design principles for shift current,
\href{https://doi.org/10.1038/ncomms14176}{Nat. Commun. \textbf{8}, 14176 (2017)}.
\bibitem{xiangThird2026}
L. Xiang, H. Jin, and J. Wang,
Third-order intrinsic anomalous Hall effect as a transport fingerprint of altermagnets,
Preprint at \href{https://doi.org/10.48550/arXiv.2604.26665}{arXiv:2604.26665 (2026)}.
\bibitem{Pizzi2020}
G. Pizzi, V. Vitale, R. Arita, S. Bl\"ugel, F. Freimuth \textit{et al.},
Wannier90 as a community code: New features and applications,
\href{https://doi.org/10.1088/1361-648X/ab51ff}{J. Phys.: Condens. Matter \textbf{32}, 165902 (2020)}.
\bibitem{Xiao2021}
J. Xiao and B. Yan,
First-principles calculations for topological quantum materials,
\href{https://doi.org/10.1038/s42254-021-00292-8}{Nat. Rev. Phys. \textbf{3}, 283 (2021)}.
\bibitem{WannierNLQG}
Z. Lu, Z. Guo, Y. Xua, J. Yao, and H. Wang,
WannierNLQG: A Julia package for nonlinear optical responses and quantum geometry from Wannier tight-binding models,
Preprint at \href{https://doi.org/10.48550/arXiv.2609.10411}{arXiv:2609.10411 (2026)}.
\bibitem{Bilbao}
S. V. Gallego, J. Etxebarria, L. Elcoro, E. S. Tasci, and J. M. Perez-Mato,
Automatic calculation of symmetry-adapted tensors in magnetic and non-magnetic materials: a new tool of the Bilbao Crystallographic Server,
\href{https://doi.org/10.1107/S2053273319001748}{Acta Crystallogr. Sect. A \textbf{75}, 438 (2019)}.
\bibitem{Liu2011PRL}
C.-C. Liu, W. Feng, and Y. Yao,
Quantum Spin Hall Effect in Silicene and Two-Dimensional Germanium,
\href{https://doi.org/10.1103/PhysRevLett.107.076802}{Phys. Rev. Lett. \textbf{107}, 076802 (2011)}.
\bibitem{Liu2011PRB}
C.-C. Liu, H. Jiang, and Y. Yao,
Low-energy effective Hamiltonian involving spin-orbit coupling in silicene and two-dimensional germanium and tin,
\href{https://doi.org/10.1103/PhysRevB.84.195430}{Phys. Rev. B \textbf{84}, 195430 (2011)}.
\bibitem{PtSe2_exp}
Y. Wang, L. Li, W. Yao, S. Song, J. T. Sun \textit{et al.},
Monolayer PtSe$_2$, a New Semiconducting Transition-Metal-Dichalcogenide, Epitaxially Grown by Direct Selenization of Pt,
\href{https://doi.org/10.1021/acs.nanolett.5b00964}{Nano Lett. \textbf{15}, 4013 (2015)}.
\bibitem{PtTe2_exp}
H. Ma, P. Chen, B. Li, J. Li, R. Ai, Z. Zhang, G. Sun, K. Yao,
Z. Lin, B. Zhao, R. Wu, X. Tang, X. Duan, and X. Duan,
Thickness-Tunable Synthesis of Ultrathin Type-II Dirac Semimetal PtTe$_2$
Single Crystals and Their Thickness-Dependent Electronic Properties,
\href{https://doi.org/10.1021/acs.nanolett.8b00583}{Nano Lett. \textbf{18}, 3523 (2018)}.
\bibitem{PtS2_exp}
J. Lu, X. Zhang, G. Su, W. Yang, K. Han, X. Yu, Y. Wan, X. Wang, and P. Yang,
Large-area uniform few-layer PtS$_2$: Synthesis, structure and physical properties,
\href{https://doi.org/10.1016/j.mtphys.2021.100376}{Mater. Today Phys. \textbf{18}, 100376 (2021)}.
\bibitem{CrI31}
M. A. McGuire, H. Dixit, V. R. Cooper, and B. C. Sales,
Coupling of crystal structure and magnetism in the layered, ferromagnetic insulator CrI$_3$,
\href{https://doi.org/10.1021/cm504242t}{Chem. Mater. \textbf{27}, 612 (2015)}.
\bibitem{CrI32}
W.-B. Zhang, Q. Qu, P. Zhu, and C.-H. Lam,
Robust intrinsic ferromagnetism and half semiconductivity in stable two-dimensional single-layer chromium trihalides,
\href{https://doi.org/10.1039/C5TC02840J}{J. Mater. Chem. C \textbf{3}, 12457 (2015)}.
\bibitem{CrI3_exp}
B. Huang, G. Clark, E. Navarro-Moratalla, D. R. Klein, R. Cheng,
K. L. Seyler, D. Zhong, E. Schmidgall, M. A. McGuire, D. H. Cobden,
W. Yao, D. Xiao, P. Jarillo-Herrero, and X. Xu,
Layer-dependent Ferromagnetism in a van der Waals Crystal down to the Monolayer Limit,
\href{https://doi.org/10.1038/nature22391}{Nature \textbf{546}, 270 (2017)}.
\bibitem{Cr2Ge2Te6_exp}
C. Gong, L. Li, Z. Li, H. Ji, A. Stern, Y. Xia, T. Cao, W. Bao,
C. Wang, Y. Wang, Z. Q. Qiu, R. J. Cava, S. G. Louie, J. Xia, and X. Zhang,
Discovery of Intrinsic Ferromagnetism in Two-Dimensional van der Waals Crystals,
\href{https://doi.org/10.1038/nature22060}{Nature \textbf{546}, 265 (2017)}.
\bibitem{FeCl2_exp}
X. Zhou, T. Jiang, Y. Tao, Y. Ji, J. Wang, T. Lai, and D. Zhong,
Evidence of Ferromagnetism and Ultrafast Dynamics of Demagnetization in an Epitaxial FeCl$_2$ Monolayer,
\href{https://doi.org/10.1021/acsnano.4c01436}{ACS Nano \textbf{18}, 10912 (2024)}.
\bibitem{Lin2016}
M. Lin, H. L. Zhuang, J. Yan, T. Z. Ward, A. A. Puretzky, C. M. Rouleau, Z. Gai, L. Liang, V. Meunier, B. G. Sumpter,
P. Ganesh, P. R. C. Kent, D. B. Geohegan, D. G. Mandrus, and K. Xiao,
Ultrathin nanosheets of CrSiTe$_3$: A semiconducting two-dimensional ferromagnetic material,
\href{https://doi.org/10.1039/C5TC03463A}{J. Mater. Chem. C \textbf{4}, 315 (2016)}.
\bibitem{Fujita2022}
R. Fujita, J. Liu, X. Hou, Y. Guo, J. Herrero-Martín, G. van der Laan, and T. Hesjedal,
X-ray spectroscopy for the magnetic study of the van der Waals ferromagnet CrSiTe$_3$ in the few- and monolayer limit,
\href{https://doi.org/10.1088/2053-1583/ac7b96}{2D Mater. \textbf{9}, 045007 (2022)}.
\bibitem{Kang2023}
S. Kang, S. Kang, H.-S. Kim, and J. Yu,
Field-controlled quantum anomalous Hall effect in electron-doped CrSiTe$_3$ monolayer,
\href{https://doi.org/10.1038/s41699-023-00375-3}{npj 2D Mater. Appl. \textbf{7}, 13 (2023)}.
\bibitem{Marrazzo2020}
A.~Marrazzo, N.~Marzari, and M.~Gibertini,
Emergent dual topology in the three-dimensional Kane-Mele
${\rm Pt}_{2}{\rm HgSe}_{3}$,
\href{https://doi.org/10.1103/PhysRevResearch.2.012063}{Phys. Rev. Research \textbf{2}, 012063(R) (2020)}.
\bibitem{footnote1}
In particular,
$j^x_{\rm inj}=2\tau( \kappa_{\rm inj;+}^{xxx} \text{Re} [E_{\perp}^x \bar{B}_{\perp}^{x*}] + \kappa_{\rm inj;+}^{xyy} \text{Re}[E_{\perp}^y \bar{B}_{\perp}^{y*}])$,
and
$j^y_{\rm inj}=2\tau( \kappa_{\rm inj;+}^{yxy} \text{Re} [E_{\perp}^x \bar{B}_{\perp}^{y*}] + \kappa_{\rm inj;+}^{yyx} \text{Re}[E_{\perp}^y \bar{B}_{\perp}^{x*}])$.

\bibitem{Burger2019}
A. M. Burger, R. Agarwal, A. Aprelev, E. Schruba, A. Gutierrez-Perez, V. M. Fridkin, and J. E. Spanier,
Direct observation of shift and ballistic photovoltaic currents,
\href{https://doi.org/10.1126/sciadv.aau5588}{Sci. Adv. \textbf{5}, eaau5588 (2019)}.
\bibitem{Li2021CIPS}
Y. Li, J. Fu, X. Mao, C. Chen, H. Liu, M. Gong, and H. Zeng,
Enhanced bulk photovoltaic effect in two-dimensional ferroelectric CuInP$_2$S$_6$,
\href{https://doi.org/10.1038/s41467-021-26200-3}{Nat. Commun. \textbf{12}, 5896 (2021)}.
\bibitem{Liang2023}
Z. Liang, X. Zhou, L. Zhang, X.-L. Yu, Y. Lv, X. Song, Y. Zhou, H. Wang,
S. Wang, T. Wang, P. P. Shum, Q. He, Y. Liu, C. Zhu, L. Wang, and X. Chen,
Strong bulk photovoltaic effect in engineered edge-embedded van der Waals structures,
\href{https://doi.org/10.1038/s41467-023-39995-0}{Nat. Commun. \textbf{14}, 4230 (2023)}.
\bibitem{xiangPtSe2}
L. Xiang, Y. Ke, and Q. Zhang,
Tunable giant Rashba-type spin splitting in PtSe$_2$/MoSe$_2$ heterostructure,
\href{https://doi.org/10.1063/1.5125303}{Appl. Phys. Lett. \textbf{115}, 203501 (2019)}.
\bibitem{Peng2026}
R.~Peng, S.~Fang, P.~Ho, F.~Liu, T.~Zhou, J.~Liu, and Y.~S.~Ang,
Ferroelastic altermagnetism,
\href{https://doi.org/10.1038/s41535-025-00835-7}{npj Quantum Mater. \textbf{11}, 5 (2026)}.
\bibitem{Gondolf2025}
J.~Gondolf, A.~Kreisel, M.~Roig, Y.~Yu, D.~F.~Agterberg, and B.~M.~Andersen,
Local signatures of altermagnetism,
\href{https://doi.org/10.1103/PhysRevB.111.174436}{Phys. Rev. B \textbf{111}, 174436 (2025)}.
\bibitem{Roig2024}
M.~Roig, A.~Kreisel, Y.~Yu, B.~M.~Andersen, and D.~F.~Agterberg,
Minimal models for altermagnetism,
\href{https://doi.org/10.1103/PhysRevB.110.144412}{Phys. Rev. B \textbf{110}, 144412 (2024)}.
\bibitem{Xu2021BSPV}
H. Xu, H. Wang, J. Zhou, and J. Li,
Pure spin photocurrent in non-centrosymmetric crystals: bulk spin photovoltaic effect,
\href{https://doi.org/10.1038/s41467-021-24541-7}{Nat. Commun. \textbf{12}, 4330 (2021)}.
\end{thebibliography}
\end{document}